\documentclass[a4]{article}

\def\beq{\begin{equation}}
\def\eeq{\end{equation}}
\def\bea{\begin{eqnarray}}
\def\eea{\end{eqnarray}}
\def\bq{\begin{quote}}
\def\eq{\end{quote}}

\def\fr{\frac}
\def\ve{\varepsilon}
\def\dt{\Delta t}
\def\dm{\Delta m}
\def\dg{\Delta \Gamma}

\def\ie{{\rm Im}(\ve)}
\def\rep{{\rm Re}(\ve)}
\def\rd{{\rm Re}(\delta)}
\def\id{{\rm Im}(\delta)}
\newcommand{\ol}{${\cal O}(\lambda^3)$}

\begin{document}

\begin{titlepage}
\begin{flushright}
FTUV-011031 \\
\end{flushright}

\vskip 1.5cm

\begin{center}
{\Large \bf      
T and CPT in B-Factories
\footnote{Talk given at the International Europhysics conference on HEP, HEP2001, 
July 2001, Budapest (Hungary)}
\par} \vskip 2.em
{\large         
{\sc J. Bernab\'eu$^{1}$, M.C. Ba\~nuls$^2$ and F. Mart\'{\i}nez-Vidal$^{3}$
}  \\[1ex] 
{\it $^1$ Depto. F\'{\i}sica Te\'orica, Universitat 
de Val\`encia (Spain)}\\ 
{\it $^2$ IFIC, Centro Mixto Universitat de Val\`encia - CSIC (Spain)} \\
{\it $^3$ INFN-Sezione di Pisa, Scuola Normale Superiore (Italy)}\\
\vskip 0.5em
\par} 

\vspace{2cm}

{\bf Abstract} \end{center} 

{For the $B_d$ meson system, CP, T and CPT indirect violation can be described
using two physical parameters, $\ve$ and $\delta$.
The traditional observables based on flavour tag and used in the kaon system,
are not helpful in the $B_d$ case, and new asymmetries have to be introduced.
Here such alternative observables, based on CP tag, are presented, together
with the first estimation on the 
sensitivity that current asymmetric B-factories can achieve on their measurement.}

\par
\vfill
\noindent

\end{titlepage}

\section{Introduction}

Violation of CP, T and CPT symmetries
in the time evolution of $K^0$-$\bar{K}^0$
was studied by the CP-LEAR experiment~\cite{cplear} 
from the preparation of definite flavour states.
The study of this \emph{flavour-to-flavour} evolution 
allows the construction of
observables which violate CP and T, or CP and CPT.
In order to be non-vanishing, nevertheless,
these observables need the presence of an absorptive part 
in the effective Hamiltonian that governs neutral meson system.
The different lifetimes of physical states $K_L$
and $K_S$ provides this ingredient.
In the case of $B_d$ mesons, on the contrary, 
the width difference $\dg$ between the physical states is expected
to be negligible\cite{kh87}, so that the T- and CPT-odd observables proposed 
for kaons, and based on flavour tag, will practically vanish for a $B_d$ system.

Here alternative observables are discussed, which allow the study of 
CP, T and CPT indirect violation in the $B_d$ system\cite{bb99.2}.
Based on CP-tag\cite{bb99}, these observables do not need the presence of $\dg \neq 0$,
and can be constructed from the entangled states of $B_d$ mesons.

In the following section, the invariant parameters $\ve$ and $\delta$ are introduced
to describe indirect violation of symmetries in the neutral meson system.
In section \ref{sec:CPtag} we describe the CP tag of $B_d$ from the entangled states
in a $B$-factory.
Next, section \ref{sec:asymmetries} reviews three different kinds of asymmetries
that can be constructed from these states, namely, \emph{flavour-to-flavour} 
and both genuine and non-genuine \emph{CP-to-flavour} asymmetries.
Finally, in section \ref{sec:experiment} the first estimates on the reach and sensitivity 
of the experimental analysis are given.

\section{Invariant description of CP, T and CPT violation in the $B$ system}
\label{sec:parameters}

The physical states in the neutral $B$-meson system are a
linear combination of the definite flavour $B^0$ and $\bar{B}^0$.
Physical states can also be written in terms of CP eigenstates, 
$|B_{\pm}\rangle \equiv \frac{1}{\sqrt{2}}(I \pm CP)|B^0\rangle$,
which are physical iff the CP operator is well defined.
To do so, one has to introduce
two complex parameters, $\ve_{1,2}$, to describe the CP mixing, so that
$|B_{1(2)} \rangle = 
\frac{1}{\sqrt{1+|\varepsilon_{1(2)}|^2}} \left [|B_{+(-)}
\rangle +\varepsilon_{1(2)}|B_{-(+)} \rangle \right ]$.
The complex parameters $\ve_{1, 2}$, invariant under rephasing of the meson states, 
are better interpreted in terms of $\ve \equiv (\ve_1+\ve_2)/2$ 
and $\delta \equiv \ve_1-\ve_2$, whose observable character is explicit
when they are written in terms of the effective hamiltonian matrix elements~\cite{bb98}.

Discrete symmetries impose different restrictions
on the effective mass matrix, $H=M-\frac{i}{2}\Gamma$,
and thus on the invariant parameters $\ve$ and $\delta$:
\begin{itemize}
\item
CPT invariance requires\footnote{Here $H_{ij}$, $M_{ij}$, and so on, represent the 
matrix elements in the flavour basis $B^0-\bar{B}^0$.} 
$H_{11}=H_{22}$, so that $\delta=0$, with no restriction on $\ve$;
\item
T invariance imposes 
${\rm Im}(M_{12} {\rm CP}_{12}^*)={\rm Im}(\Gamma_{12}{\rm CP}_{12}^*)=0$,
and so $\ve=0$;
\item
and CP conservation requires both $\ve=\delta=0$.
\end{itemize}
In the exact limit $\Delta \Gamma=0$, an approximation that is 
expected to be excellent for the $B_d$ system, 
both ${\rm Re}(\ve)$ and ${\rm Im}(\delta)$ vanish.
Then $\ie \neq 0$ is a proof of both CP and T violation, and $\rd \neq 0$
is a proof of CP and CPT violation, but neither $\rep=0$ nor $\id =0$
are proof of a fundamental invariance.
Information on the symmetry parameters can be extracted 
from the study of time evolution of $B$ meson entangled states.

\section{CP-Tag from entangled states}
\label{sec:CPtag}

In a $B$ factory operating at the $\Upsilon(4S)$ peak, 
correlated pairs of neutral $B$-mesons are produced through
$e^+ e^- \rightarrow \Upsilon(4S) \rightarrow B \bar{B}$.
Charge conjugation together with Bose statistics require
the initial state to be
\beq
\vert i>=\frac{1}{\sqrt{2}} 
\left (\vert B^0(\vec{k}), \overline{B}^0(-\vec{k})>
- \vert \overline{B}^0(\vec{k}), B^0(-\vec{k})> \right ).
\label{eq:ent}
\eeq
This permits the performance of a flavour tag: if at $t_0$ 
one of the mesons decays through a channel $X$, which is only allowed for one flavour,
the other meson in the pair must have the opposite flavour at $t_0$,
and will later evolve during $\dt=t-t_0$ until its final decay to some state $Y$.

The entangled $B-\bar{B}$ state can also be expressed 
in terms of the CP eigenstates as
$\vert i>=\frac{1}{\sqrt{2}} \left (\vert B_-, B_+>
- \vert B_+, B_-> \right )$. 
Thus it is also possible to carry out a CP tag, 
if we have a CP-conserving decay into a definite CP final state $X$, 
so that its detection allows us to identify the decaying meson 
as a $B_+$ or a $B_-$, which decays into $Y$ after a time $\dt$.
In Ref.~\cite{bb99} we described how this determination is possible
and unambiguous to \ol, the flavour-mixing parameter of the CKM matrix.

If we consider only decay channels $X$, $Y$ which are either 
flavour or CP conserving,
then the final configuration $(X,\,Y)$ corresponds to a single particle 
mesonic transition.
The intensity for the final configuration, $I(X,\,Y,\dt)\equiv 
\fr{1}{2}\int_{\dt}^{\infty} dt' |(X,\,Y)|^2$
is proportional to the time dependent probability for the meson transition.

\section{Asymmetries}
\label{sec:asymmetries}

By comparing the probabilities corresponding to different processes 
we build time-dependent asymmetries that 
can be classified into three types.

\subsection{Flavour-to-flavour genuine asymmetries}
The final configuration denoted by $(\ell,\ell)$, with flavour definite
(for example, semileptonic) 
decays detected on both sides of the detector, corresponds to
\emph{flavour-to-flavour} transition at the meson level.
The equivalence is shown in Table~\ref{tab:f2f}.
\begin{table}[htb]
\caption{
\emph{Flavour-to-flavour} transitions.}
\label{tab:f2f}
\begin{center}
\begin{tabular}{cc}
\hline
$(X,\, Y)$ & Meson Transition \\
\hline 
$(\ell^+, \, \ell^+)$ & $\bar{B}^0 \stackrel{\phantom{c}}{\rightarrow} {B^0}$ \\
$(\ell^-, \, \ell^-)$ & $B^0 \rightarrow \bar{B}^0$ \\
$(\ell^+, \, \ell^-)$ & $\bar{B}^0 \rightarrow \bar{B}^0$ \\
$(\ell^-, \, \ell^+)$ & $B^0 \rightarrow B^0$ \\
\hline
\end{tabular}
\end{center}
\end{table}
The first two processes in the Table
are conjugated under CP and also under T. The corresponding Kabir
asymmetry\cite{ka68} is, to linear order in the CPT violating $\delta$,
\beq
A (\ell^+, \, \ell^+) \approx 
{\scriptstyle \fr{4 \fr{{\rm Re}(\ve)}{1+ |\ve|^2}}{1+4 \fr{{\rm Re}(\ve)}{1+ |\ve|^2}}},
\label{eq:l+l+}
\eeq
which does not depend on time.
However, in the exact limit $\dg=0$, $\rep$ vanishes,
and this quantity will be zero.
For the $B_d$ system, experimental limits on $\rep$ are of few parts in a thousand\cite{ac97}
\cite{bbar01}.

A second asymmetry arises from the last two processes in Table~\ref{tab:f2f},
related by a CP or a CPT transformation,
\beq
A(\ell^+ , \, \ell^-) \approx 
- 2
\fr{
{\rm Re} \!\left (\! {\scriptstyle\frac{\delta}{1-\varepsilon^2}}\!\right ) {\sinh} {\scriptstyle\frac{\Delta \Gamma \Delta t}{2}} 
\!-\! {\rm Im} \!{\scriptstyle\left ( \!\frac{\delta}{1-\varepsilon^2}\!\right )} \sin({\scriptstyle \Delta m \Delta t})}
{{\cosh} {\scriptstyle\frac{\dg \dt}{2}}\!+\! \cos({\scriptstyle\Delta m \dt})},
\label{eq:l+l-}
\eeq
which is an odd function of time.
This asymmetry also vanishes unless $\dg \neq 0$.
Present limits~\cite{ac97} on $\id$ are at the level of few percent.

\subsection{CP-to-flavour genuine asymmetries}

Alternative asymmetries can be constructed
making use of the CP eigenstates, 
which can be identified in this system by means of a CP tag.
If the first decay product, $X$, is a CP eigenstate
produced along the CP-conserving direction, i.e. the decay is free of CP violation, 
and $Y$ is a flavour definite channel, then the mesonic transition corresponding
to the configuration $(X, \, Y)$ is of the type \emph{CP-to-flavour}.

In Table~\ref{tab:CP2f} we show the mesonic transitions, with their 
related final configurations, connected by genuine symmetry transformations
to $B_+ \rightarrow B^0$.
\begin{table}[htb]
\caption{
Transitions connected to $(J/\Psi K_S, \, \ell^+)$.}
\label{tab:CP2f}
\begin{center}
\begin{tabular}{ccc}
\hline
$(X,\, Y)$ & Transition & Transformation \\ 
\hline
$(J/\Psi K_S, \, \ell^-)$ & $B_+ \rightarrow \bar{B}^0$ & CP\\
$(\ell^-, \, J/\Psi K_L)$ & $B^0 \rightarrow B_+$ & T\\
$(\ell^+, \, J/\Psi K_L)$ & $\bar{B}^0 \rightarrow B_+$ & CPT\\
\hline
\end{tabular}
\end{center}
\end{table}

Comparing the intensities of the four processes,
we may construct three genuine asymmetries, namely $A({\rm CP})$,
$A({\rm T})$ and $A({\rm CPT})$ \cite{bb99.2}.
\beq
A({\rm CP}) =
-2 {\scriptstyle \fr{{\rm Im}(\ve)}{1+|\ve|^2}} \sin ({\scriptstyle \dm \dt}) 
+ {\scriptstyle \fr{1-|\ve|^2}{1+|\ve|^2}} 
{\scriptstyle \fr{2 {\rm Re}(\delta)}{1+|\ve|^2}} 
\sin^2 \left ({\scriptstyle \fr{\dm\dt}{2}}\right ),
\label{eq:aCP}
\eeq
the CP odd asymmetry,
contains both T-violating and CPT-violating contributions,
which are, respectively, odd and even functions of $\dt$.
This asymmetry corresponds to the "gold plate" decay \cite{bigi}
and has been measured recently \cite{abe00}.
T and CPT violating terms can be separated by constructing other 
asymmetries.
\beq
A({\rm T}) =
-2 {\scriptstyle \fr{{\rm Im}(\ve)}{1+|\ve|^2}} \sin ({\scriptstyle\dm \dt}) 
 \left[ 1 - {\scriptstyle \fr{1-|\ve|^2}{1+|\ve|^2}} 
{\scriptstyle \fr{2 {\rm Re}(\delta)}{1+|\ve|^2}} 
\sin^2 \left ({\scriptstyle \fr{\dm \dt}{2}}\right) \right],
\label{eq:aT}
\eeq
the T asymmetry, 
needs $\ve \neq 0$, and turns out to be purely odd in $\dt$ in the limit we are considering. 
\beq
A({\rm CPT})=
{\scriptstyle \fr{1-|\ve|^2}{1+|\ve|^2}} 
{\scriptstyle \fr{2 {\rm Re}(\delta)}{1+|\ve|^2}} 
\fr{\sin^2 \left ({\scriptstyle\fr{\dm \dt}{2}}\right )}
{1-2 \fr{{\rm Im}(\ve)}{1+|\ve|^2} \sin ({\scriptstyle\dm \dt})},
\label{eq:aCPT}
\eeq
is the CPT asymmetry. It
needs $\delta \neq 0$, and includes both even and odd time dependences.

The above expressions correspond to the limit $\dg=0$, but,
being genuine observables,
 a possible absorptive part could not
induce by itself a non-vanishing asymmetry.

\subsection{CP-to-flavour non-genuine asymmetries}

The construction of the quantities described in the previous paragraphs
requires to tag both $B_+$ and $B_-$ states, and thus the reconstruction
of the experimentally challenging decay $B \rightarrow J/\Psi K_L$.
Conversely, non-genuine asymmetries offer a possibility to measure the symmetry 
parameters from the reconstruction of $J/\Psi K_S$ only.
But they involve the discrete transformation that we denote $\dt$, consisting of the
exchange in the order of appearance of decay products $X$ and $Y$,
which cannot be associated with any fundamental symmetry.

Table~\ref{tab:CP2fng} shows the different transitions 
we may study from such final states.
\begin{table}[htb]
\caption{
Final configurations with only $J/\Psi K_S$.}
\label{tab:CP2fng}
\begin{center}
\begin{tabular}{ccc}
\hline
$(X,\, Y)$ & Transition & Transformation \\ 
\hline
$(J/\Psi K_S, \, \ell^-)$ & $B_+ \rightarrow B^0$ & CP\\
$(\ell^+, \, J/\Psi K_S)$ & $\bar{B}^0 \rightarrow B_-$ & $\dt$\\
$(\ell^-, \, J/\Psi K_S)$ & $\bar{B}^0 \rightarrow B_-$ & $\dt$+CP\\
\hline
\end{tabular}
\end{center}
\end{table}
Besides the genuine CP asymmetry, there are two new quantities
that can be constructed from
the comparison between $(J/\Psi K_S, \, \ell^+)$ and the
processes in the table.
In the exact limit $\dg =0$, $\dt$ and T operations are found to
become equivalent, so that the temporal asymmetry satisfies
$A(\dt)\equiv A(\ell^+, \, J/\Psi K_S) = A({\rm T})$ and moreover
$A({\rm CP}\dt) \equiv A(\ell^-, \, J/\Psi K_S) = A({\rm CPT})$.
Since this result holds for $\dg\approx0$, it is expected to be valid 
for the $B_d$ system, but not for $B_s$ and even less for $K$.
The asymmetries $A(\dt)$ and $A({\rm CP}\dt)$ are non-genuine, and
the presence of $\dg \neq 0$ may induce
non-vanishing values for them, 
even if there is no true T or CPT violation.
These fake effects, nevertheless, can be calculated and are thus controllable.

\section{CP, T, CPT indirect violation reach at asymmetric B-Factories}
\label{sec:experiment}

The asymmetries described in the previous section can be already constructed 
from the current data taken at Asymmetric B-Factories \cite{abe00}.
The experimental analysis is based on a simultaneous unbinned likelihood fit 
of the flavour and CP intensities $I(X,\,Y;\dt)$,
together with the $B^0$/$\bar{B}^0$ mistag rates
and the $\dt$ resolution function.
The coefficients of terms with different temporal dependencies contain the 
information on the symmetry parameters.

\begin{table}[htb]
\caption{Projections for 60 fb$^{-1}$.}
\label{tab:exp}
\begin{center}
\begin{tabular}{lcc}
\hline
Parameter & (Generated) & Statistical error \\
\hline
$\fr{\rd}{1+|\ve|^2}$ & (0)& 0.09 \\
$\fr{\rep}{1+|\ve|^2}$ & (0) & 0.007 \\
$\fr{\dg}{\Gamma}$ & (0) & 0.07 \\
$\fr{\ie}{1+|\ve|^2}$ & (0.35)& 0.04 \\
$\dm$ & (0.472 ps$^{-1}$) & 0.009 \\
\hline
\end{tabular}
\end{center}
\end{table}

From a simulation study, estimations on the reachable statistical 
precision for the relevant parameters
have been calculated for $\approx$60 fb$^{-1}$ (assuming yields
from Ref. \cite{abe00}) and are shown in Table~\ref{tab:exp}.

\section{Conclusions}

We have shown how the two complex rephasing invariant parameters $\ve$ and $\delta$
describe CP, T and CPT indirect violation in $B^0-\bar{B}^0$.
In the exact limit $\dg=0$ the number of parameters is reduced to $\ie$
and $\rd$.
Observables based on  
{\it flavour-to-flavour} transitions are sensitive to $\rep$,
but need $\dg \! \neq \! 0$, and thus are not promising in B-factories.
Conversely, these experimental facilities allow the construction of new 
asymmetries based on combination of flavour and CP tags.

First estimations on the sensitivity reachable on B-factories
have been presented.
This data will be crucial to achieve the separation
of the two ingredients: on one hand CP and T violation, described by $\ve$, and
on the other CP and CPT violation, given by $\delta$.

\vspace*{.5 cm}

This work has been supported by CICYT, Spain, under Grant AEN99-0692.


\end{document}